\documentclass[prb,showpacs,twocolumn,superscriptaddress,aps,a4paper]{revtex4}
\usepackage{dcolumn}
\usepackage{amsmath}
\usepackage{graphicx}
\usepackage{latexsym}
\usepackage{amsfonts}
\usepackage{amssymb}
\DeclareGraphicsExtensions{.pdf,.gif,.jpg}

\newcommand{\be}{\begin{equation}}
\newcommand{\ee}{\end{equation}}
\newcommand{\beq}{\begin{eqnarray}}
\newcommand{\eeq}{\end{eqnarray}}

\begin{document}

\def\bbe{\mbox{\boldmath $e$}}
\def\bbf{\mbox{\boldmath $f$}}
\def\bg{\mbox{\boldmath $g$}}
\def\bh{\mbox{\boldmath $h$}}
\def\bj{\mbox{\boldmath $j$}}
\def\bq{\mbox{\boldmath $q$}}
\def\bp{\mbox{\boldmath $p$}}
\def\br{\mbox{\boldmath $r$}}
\def\bz{\mbox{\boldmath $z$}}

\def\bfzero{\mbox{\boldmath $0$}}
\def\bfone{\mbox{\boldmath $1$}}

\def\dr{{\rm d}}

\def\tb{\bar{t}}
\def\zb{\bar{z}}

\def\tgb{\bar{\tau}}

\def\bC{\mbox{\boldmath $C$}}
\def\bG{\mbox{\boldmath $G$}}
\def\bH{\mbox{\boldmath $H$}}
\def\bK{\mbox{\boldmath $K$}}
\def\bM{\mbox{\boldmath $M$}}
\def\bN{\mbox{\boldmath $N$}}
\def\bO{\mbox{\boldmath $O$}}
\def\bQ{\mbox{\boldmath $Q$}}
\def\bR{\mbox{\boldmath $R$}}
\def\bS{\mbox{\boldmath $S$}}
\def\bT{\mbox{\boldmath $T$}}
\def\bU{\mbox{\boldmath $U$}}
\def\bV{\mbox{\boldmath $V$}}
\def\bZ{\mbox{\boldmath $Z$}}

\def\bcalS{\mbox{\boldmath $\mathcal{S}$}}
\def\bcalG{\mbox{\boldmath $\mathcal{G}$}}
\def\bcalE{\mbox{\boldmath $\mathcal{E}$}}

\def\bgG{\mbox{\boldmath $\Gamma$}}
\def\bgL{\mbox{\boldmath $\Lambda$}}
\def\bgS{\mbox{\boldmath $\Sigma$}}

\def\bgr{\mbox{\boldmath $\rho$}}
\def\bgs{\mbox{\boldmath $\sigma$}}

\def\a{\alpha}
\def\b{\beta}
\def\g{\gamma}
\def\G{\Gamma}
\def\d{\delta}
\def\D{\Delta}
\def\e{\epsilon}
\def\ve{\varepsilon}
\def\z{\zeta}
\def\h{\eta}
\def\th{\theta}
\def\k{\kappa}
\def\l{\lambda}
\def\L{\Lambda}
\def\m{\mu}
\def\n{\nu}
\def\x{\xi}
\def\X{\Xi}
\def\p{\pi}
\def\P{\Pi}
\def\r{\rho}
\def\s{\sigma}
\def\S{\Sigma}
\def\t{\tau}
\def\f{\phi}
\def\vf{\varphi}
\def\F{\Phi}
\def\c{\chi}
\def\w{\omega}
\def\W{\Omega}
\def\Q{\Psi}
\def\q{\psi}

\def\ua{\uparrow}
\def\da{\downarrow}
\def\de{\partial}
\def\inf{\infty}
\def\ra{\rightarrow}
\def\bra{\langle}
\def\ket{\rangle}
\def\grad{\mbox{\boldmath $\nabla$}}
\def\Tr{{\rm Tr}}
\def\hc{{\rm h.c.}}

\title{Spin-flip scattering in time-dependent transport through a quantum
dot:\\
Enhanced spin-current and inverse tunneling magnetoresistance}


\author{Enrico Perfetto}
\affiliation{Dipartimento di Scienza dei Materiali, Universit\'a
di Milano-Bicocca, Via Cozzi 53, 20125 Milano,
Italy}
\affiliation{Laboratori Nazionali di Frascati, Istituto
Nazionale di Fisica Nucleare, Via E. Fermi 40, 00044 Frascati,
Italy}

\author{Gianluca Stefanucci}
\affiliation{Dipartimento di Fisica, Universit\'a di Roma Tor
Vergata, Via della Ricerca Scientifica 1, I-00133 Rome, Italy}
\affiliation{European Theoretical Spectroscopy Facility (ETSF)}
\affiliation{Laboratori Nazionali di Frascati, Istituto Nazionale
di Fisica Nucleare, Via E. Fermi 40, 00044 Frascati, Italy}

\author{Michele Cini}
\affiliation{Dipartimento di Fisica, Universit\'a di Roma Tor
Vergata, Via della Ricerca Scientifica 1, I-00133 Rome, Italy}
\affiliation{Laboratori Nazionali di Frascati, Istituto Nazionale
di Fisica Nucleare, Via E. Fermi 40, 00044 Frascati, Italy}

\date{\today}

\begin{abstract}

We study the effects of spin-flip scatterings on the
time-dependent transport properties through a magnetic quantum dot
attached to normal and ferromagnetic leads. The transient
spin-dynamics as well as the steady-state tunneling
magnetoresistance (TMR) of the system are investigated. The
absence of a definite spin quantization axis requires the
time-propagation of two-component spinors. We present numerical
results in which the electrodes are treated both as
one-dimensional tight-binding wires and in the wide-band limit
approximation. In the latter case we derive a transparent analytic
formula for the spin-resolved current, and transient oscillations
damped over different time-scales are identified. We also find a
novel regime for the TMR inversion. For any given strength of the
spin-flip coupling the TMR becomes negative provided the
ferromagnetic polarization is larger than some critical value.
Finally we show how the full knowledge of the transient response
allows for enhancing the spin-current by properly tuning the
period of a pulsed bias.

\end{abstract}



\maketitle


\section{Introduction}
\label{intro}

Single-electron spin in nanoscale systems is a promising building
block for both processors and memory storage devices in future
spin-logic applications.\cite{loss1} In particular great attention
has been given to quantum dots (QDs), which are among the best
candidates to implement quantum bit gates for quantum
computation.\cite{loss} In these systems the spin coherence time
can be orders of magnitude longer than the charge coherence
times,\cite{time1,time2,time3,time4} a feature which allows one to
perform a large number of operations before the spin-coherence is
lost. Another advantage of using QDs is the possibility of
manipulating their electronic spectrum with, e.g., external
magnetic fields and gate voltages, and fine-tuning the
characteristic time-scales of the system.

For practical applications, it is important to achieve full
control of the ultrafast dynamics of QD systems after the sudden
switch-on of an external perturbation. To this end the study of
the time evolution of spin-polarized currents and the
characterization of the spin-decoherence time is crucial. The
theoretical study of the transient regime is also relevant in the
light of recent progresses in the time-resolution of dynamical
responses at the nanoscale. Novel techniques like transient
current spectroscopy\cite{transient} and time-resolved Faraday
rotation\cite{transient2,transient3} allow us to follow the
microscopic dynamics at the sub-picosecond time-scale. These
advances may open completely new scenarios with respect to those
at the steady-state. For instance, transient coherent quantum
beats of the spin dynamics in semiconductor QD's have been
observed after a circularly polarized optical
excitation,\cite{transient2,transient3} and theoretically
addressed by Souza.\cite{souza} We also mention a recent
experiment on split-gate quantum point contacts in which the
measured quantum capacitance in the transient regime was six
orders of magnitude larger than in the
steady-state.\cite{transient4}

Another attracting feature of magnetic QDs is their use as
spin-devices in magnetic tunnel junctions (MTJs). Different
orientations of the polarization of ferromagnetic leads results in
spin-dependent tunneling rates and hence in a nonvanishing
tunneling magnetoresistance (TMR). In recent
experiments\cite{tmrinversion,tmrinversion2,tmrinversion3} the
inverse TMR effect (TMR $<0$) has been observed and various models
have been proposed to address such
phenomenon.\cite{tsymbal,tmrneg1,tmrneg2}

In this paper we study the time-dependent transport through a
single level quantum dot connected to normal and ferromagnetic
leads. In order to get a sensible transient regime, we adopt the
so-called partition-free approach, in which the
electrode-QD-electrode system is assumed to be in equilibrium
before the external bias is switched on.\cite{cini,stefanucci}
Spin-symmetry breaking terms (like spin-flip scatterings and
spin-dependent dot-leads hoppings) are included\cite{souza3} and
this requires the time-propagation of a genuine two-component
spinor.

Explicit calculations are performed both in the case of
non-interacting one-dimensional (1D) leads of finite length and in
the wide-band-limit approximation  (WBLA). In the WBLA we derive a
closed analytic formula for the exact spin-resolved time-dependent
current, which can be expressed as the sum of a steady-state and a
transient contribution. The latter consists of a term which only
contains transitions between the QD levels and the electrochemical
potentials (\textit{resonant-continuum}), and a term which only
contains intra-dot transitions (\textit{resonant-resonant}).
Remarkably these two terms are damped over two different
time-scales, with the resonant-continuum transitions longer lived
than the resonant-resonant transitions. We further show that,
going beyond the WBLA, extra transitions occur. These involve the
top/bottom of the 1D electrode bands and might be relevant for
narrow-band electrodes.

For QDs connected to normal leads we study the quantum beat
phenomenon in the presence of intra-dot spin-flip coupling of
strength $V_{sf}$. We show that the amplitude of the beats is
suppressed independently of the structure of the leads (WBLA and
1D leads). In addition we show how to engineer the
spin-polarization of the total current by exploiting the full
knowledge of the time-dependent response of the system. By
applying a periodic pulse of proper period, we provide numerical
evidence of oscillating spin-polarizations with amplitude two
orders of magnitude larger than in the DC case.

Finally, in the case of ferromagnetic electrodes we study the
steady-state TMR and the conditions for its inversion. Treating
the leads in the WBLA we derive a simple formula for the TMR in
linear response. Noticeably, in the presence of spin flip
interactions there is a critical value of the polarization of the
ferromagnetic leads above which the TMR is negative even for
symmetric contacts to the left/right leads. This provides an
alternative mechanism for the TMR inversion. The above scenario is
qualitatively different with 1D leads, e.g., the TMR can be
negative already for $V_{sf}=0$.

The paper is organized as follows. In Section \ref{sec1} we
introduce the lead-QD-lead model. In Section \ref{sec2} we employ
the WBLA and derive an exact formula for the time-dependent
current in the presence of spin-symmetry breaking terms. We
present explicit results both for normal and ferromagnetic leads,
focussing on transient quantum beats (normal) and the TMR
(ferromagnetic). In Section \ref{sec3} we address the above
phenomena in the case of 1D leads and engineer the dynamical spin
responses. Here we also describe the numerical framework employed
to compute the time-dependent evolution of the system. The summary
and main conclusions are drawn in Section \ref{secV}.

\section{Model}
\label{sec1}

\begin{figure}[htbp]
    \vspace{0.2cm}
\includegraphics*[width=.47\textwidth,height=2.5cm]{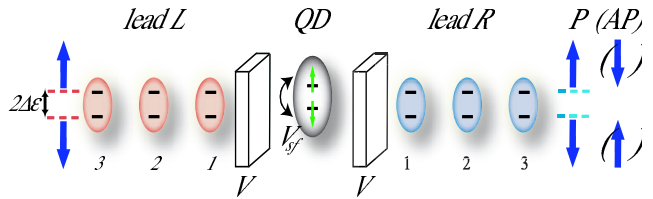}
\caption{(Colour online) Schematic illustration of the model
Hamiltonian. The lead $L$ with majority spin $\ua$ electrons and
the lead $R$ with majority spin $\ua$ ($\da$) electrons in the P
(AP) configurations are separated from the QD by tunnel barriers
which are accounted for with a renormalized hopping $V$. The
energy-spin-splitting is $2\Delta \varepsilon$ in the leads and
$2E_{z}$ in the QD. Spin-flip scatterings of strength $V_{sf}$
occur in the QD.} \label{device}
\end{figure}

We consider the system illustrated in Fig.\ref{device} which
consists of a single-level QD contacted with 1D Left ($L$) and
Right ($R$) electrodes. The latter are described by the
tight-binding Hamiltonian
\begin{eqnarray}
H_{\alpha}&=& \sum_{\sigma =\uparrow,\downarrow}\sum_{j=1}^{N}
\left( \varepsilon_{\alpha j \sigma} c^{\dagger}_{\alpha j
\sigma} c_{\alpha j \sigma} \right. \nonumber \\
&+& \left.  V_{0}c^{\dagger}_{\alpha j \sigma} c_{\alpha j+1
\sigma} + V_{0}^{*}c^{\dagger}_{\alpha j+1 \sigma} c_{\alpha j
\sigma} \right) \, , \label{leads}
\end{eqnarray}
where $N$ is the number of sites of lead $\alpha = L,R$, the
nearest-neighbor hopping integral is $V_{0}$ and
$c^{(\dagger)}_{\alpha i \sigma}$ is the annihilation (creation)
operator of an electron on site $i$ of the lead $\alpha$ with spin
$\sigma$. In the case of ferromagnetic leads, we distinguish
between two configurations, one with parallel (P) and the other
with antiparallel (AP) magnetization of the two leads. In order to
model these two different alignments we use the Stoner
prescription according to which the on-site energies are

\begin{eqnarray}
\varepsilon_{\alpha j \sigma} &=& (-1)^{\delta_{\sigma,
\downarrow}} \Delta \varepsilon \quad\quad\quad\quad\quad\quad
\mathrm{P \quad } \nonumber \\
\varepsilon_{\alpha j \sigma} &=&
(-1)^{\delta_{\alpha,L}}(-1)^{\delta_{\sigma, \downarrow}} \Delta
\varepsilon \quad\quad\; \mathrm{AP \quad }
\end{eqnarray}
where $2 \Delta \varepsilon$ is the band spin splitting. In the
above model  the P configuration corresponds to a majority-spin
electron with spin $\uparrow$ in both leads, while in the AP
configuration the majority-spin electron have spin $\uparrow$ in
lead $L$ and spin $\downarrow$ in lead $R$. The case of normal
leads corresponds to $\Delta \varepsilon =0$. The Hamiltonian of
the quantum dot reads
\begin{equation}
H_{QD}= \sum_{\sigma} \varepsilon_{d \sigma}
d^{\dagger}_{\sigma}d_{\sigma}+V_{sf}d^{\dagger}_{\uparrow}d_{\downarrow}
+V_{sf}^{*}d^{\dagger}_{\downarrow}d_{\uparrow} \, ,
\end{equation}
where $d^{(\dagger)}_{\sigma}$  annihilates (creates) an electron
on the QD with spin $\sigma$ and $V_{sf}$ is responsible for
intra-dot spin-flip scatterings. The on-site energy
$\varepsilon_{d \sigma}=\varepsilon_{d}+\sigma E_{z}/2$, $\sigma =
\pm 1$, where $E_{z}$ is an intra-dot energy splitting. The
quantum dot is connected to the leads via the tunneling
Hamiltonian
\begin{equation}
H_{T}=\sum_{\alpha=L,R}\sum_{\sigma \sigma'} \left(V_{d\sigma
,\alpha \sigma'} d^{\dagger}_{\sigma}c_{\alpha 1 \sigma'} + V_{
d\sigma, \alpha \sigma'}^{*} c^{\dagger}_{\alpha 1
\sigma'}d_{\sigma} \right) \, . \label{ht}
\end{equation}
with $V_{d\sigma ,\alpha \sigma'}$ the amplitude for an electron
in the QD with spin $\sigma$ to hop to the first site of lead
$\alpha$ with spin $\sigma'$. Alternatively, one can express
$H_{T}$ in terms of the one-body eigenstates of $H_{\alpha}$
labelled by $(\alpha, k, \sigma)$,
\begin{equation}
H_{T}=\sum_{\alpha=L,R} \sum_{k=1}^{N} \sum_{\sigma \sigma'}
\left(V_{d\sigma, \alpha k \sigma'} d^{\dagger}_{\sigma}c_{\alpha
k \sigma'} + V_{d\sigma,\alpha k \sigma'}^{*} c^{\dagger}_{\alpha
k \sigma'}d_{\sigma} \right) \, ,
\end{equation}
with $V_{d\sigma, k\alpha \sigma'}=\sqrt{2/(N+1)} \,  V_{d\sigma
,\alpha \sigma'} \, \sin k $.

Putting all terms together the Hamiltonian of the whole system in
equilibrium reads
\begin{equation}
H_{0}=H_{L}+H_{R}+H_{QD}+H_{T} \, .
\label{htot}
\end{equation}

In the next two Sections we study out-of-equilibrium properties of
the system described in Eq.(\ref{htot}), after a sudden switch-on
of a bias voltage $U_{\alpha}(t)$ in lead $\alpha=L,R$. We
calculate the time-dependent spin-polarized current $I_{\alpha
\sigma}(t)$ flowing between the QD and lead $\alpha$. $I_{\alpha
\sigma}(t)$ is defined as the variation of the total number of
particles $N_{\alpha \sigma}$ of spin $\sigma$ in lead $\alpha$,
\begin{equation}
I_{\alpha \sigma}(t) \equiv \frac{d}{dt}N_{\alpha \sigma} = 2\Re
\sum_{k,\sigma'} V_{\alpha k \sigma,d \sigma'} \, G^{<}_{d
\sigma',\alpha k \sigma}(t,t) \, , \label{current}
\end{equation}
where $G^{<}_{d \sigma',\alpha k \sigma}$ is the dot-lead lesser
Green's function of the contacted non-equilibrium system, and
$\Re$ stands for the real part. Unless otherwise stated, the
current is computed at the $L$ interface and the short-hand
notation $I_{\sigma}=I_{L \sigma}$ is used for the spin-polarized
current while $I_{tot}=I_{\uparrow}+I_{\downarrow}$ and
$I_{spin}=I_{\uparrow}-I_{\downarrow}$ denote the total current
and spin current respectively. We further specialize to constant
biases $U_{\alpha}(t)=U_{\alpha}$, except in Section \ref{ivb}
where the bias is a periodic pulse. Therefore we define $U \equiv
U_{L}$, while the value of $U_{R}$ is either 0 or $-U$. The
numerical simulations have been performed using dot-leads hoppings
independent of $\alpha$ (symmetric contacts). We show that,
despite the symmetry between $L$ and $R$ leads, the sign of the
steady-state TMR can be inverted, a property which has been so far
ascribed to strongly asymmetric tunneling
barriers.\cite{tsymbal,rlxh.2007} Moreover we specialize
Eq.(\ref{ht}) to the case $V_{d\uparrow, \alpha \uparrow}=V_{d
\downarrow, \alpha \downarrow} \equiv V$, and $V_{d \uparrow,
\alpha \downarrow}=V_{d \downarrow, \alpha \uparrow}=0$. However
we note that spin-flip scatterings are accounted for by
considering a finite intra-dot $V_{sf}$ which is taken to be real
in this work. If not otherwise specified, the initial Fermi energy
$E_{F}$ of both leads is set to 0 (half-filled electrodes). In the
numerical calculations below all energies are measured in units of
$V_{0}$ ($4V_{0}$ is the bandwidth of the leads), times are
measured in units of $\hbar/V_{0}$ and currents in units of $e
V_{0}/\hbar$ with $e$ the electron charge.

\section{Results in the wide band limit approximation}
\label{sec2}

Here we study the model introduced in the previous Section within
the WBLA. This approximation has been mostly used to study
spinless electrons and a closed formula for the exact
time-dependent current has been derived.\cite{jahuo,stefanucci}
Below we generalize these results to systems where the spin
symmetry is broken. The retarded Green's function projected onto
the QD can be expressed in terms of the embedding self-energy
\begin{equation}
\Big{(}\Sigma (\omega) \Big{)}_{\sigma \sigma'}= \sum_{\nu =
\uparrow,\downarrow} \,  \sum_{\alpha = L,R} \Big{(}\Sigma_{\alpha
\nu} (\omega) \Big{)}_{\sigma \sigma'} \label{self}
\end{equation}
which is a $2 \times 2$ matrix in spin space. In Eq.(\ref{self})
$(\Sigma_{\alpha \nu})_{\sigma \sigma'}$ accounts for virtual
processes where an electron  on the QD hops to lead $\alpha$ by
changing its spin from $\sigma$  to $\nu$, and hops back to the QD
with final spin $\sigma'$. Exploiting the Dyson equation the
expression of $(\Sigma_{\alpha \nu})_{\sigma \sigma'}$ is
\begin{equation}
\Big{(}\Sigma_{\alpha \nu} (\omega) \Big{)}_{\sigma
\sigma'}=\sum_{k} \, V_{d\sigma, \alpha k \nu} \, g_{\alpha k \nu}
(\omega) \, V_{\alpha k \nu, d\sigma'} \, , \label{selfexp}
\end{equation}
where $ g_{\alpha k \nu} = (\omega -\varepsilon_{\alpha k
\nu}-U_{\alpha} +i \eta)^{-1}$ is the retarded Green's function of
the uncontacted $\alpha$ lead expressed in term of the
one-particle eigenenergies $\varepsilon_{\alpha k \nu}$. In the
WBLA Eq.(\ref{selfexp}) becomes
\begin{equation}
\Big{(}\Sigma_{\alpha \nu} (\omega) \Big{)}_{\sigma \sigma'}
\approx -\frac{i}{2} \Big{(} \Gamma_{\alpha \nu} \Big{)}_{\sigma
\sigma'}  \, ,\label{wbl}
\end{equation}
i.e., the self-energy is independent of frequency. The $2 \times
2$ matrices $ \Gamma$'s have the physical meaning of
spin-dependent tunneling rates and account for spin-flip processes
between the leads and the QD. We wish to point out that each
$\Gamma$ matrix must be positive semi-definite for a proper
modeling of the WBLA. Indeed, given an arbitrary two dimensional
vector $(v_{\uparrow},v_{\downarrow})$ one finds
\begin{equation}
\sum_{\sigma , \sigma'} v^{*}_{\sigma} \Big{(} \Gamma_{\alpha \nu}
\Big{)}_{\sigma \sigma'} v_{\sigma'} = 2 \pi \sum_{k}
\delta(\omega - \varepsilon_{\alpha k \nu}-U_{\alpha})
\Big{|}\sum_{\sigma}V_{d \sigma , \alpha k \nu}
v_{\sigma}\Big{|}^{2} \,.
\end{equation}
The above condition ensures the damping of all transient effects
in the calculation of local physical observables.

Proceeding along similar lines as in Ref.\onlinecite{stefanucci}
one can derive an explicit expression for spin-polarized current
$I_{\alpha \sigma}(t)$ defined in Eq.(\ref{current}),
\begin{equation}
I_{\alpha \sigma}(t)= I_{\alpha \sigma}^{s}+ \int \frac{d\omega}{2
\pi} \, f(\omega) \, \mathrm{Tr} \Big{[} T_{\alpha}(\omega,t)
\Gamma_{\alpha \sigma} \Big{]} \, , \label{itot}
\end{equation}
$f(\omega)$ being the Fermi distribution function. In the above
equation $I_{\alpha \sigma}^{s}$ is the steady-state polarized
current which for $\alpha =L$ reads
\begin{eqnarray}
I_{L \sigma}^{s} &=& \int \frac{d\omega}{2 \pi} \,
\Big{(}f(\omega-U_{L})-f(\omega -U_{R}) \Big{)}
 \nonumber \\ &\times& \mathrm{Tr}
\Big{[} \frac{1}{\omega-H_{QD}+\frac{i}{2}\Gamma} \Gamma_{R}
\frac{1}{\omega-H_{QD}-\frac{i}{2}\Gamma}  \Gamma_{L \sigma}
\Big{]} \nonumber \\
\label{ssl}
\end{eqnarray}
with $\Gamma_{\alpha}=\sum_{\sigma} \Gamma_{\alpha \sigma}$ and
$\Gamma=\sum_{\alpha} \Gamma_{\alpha}$. The steady current at the
right interface $I_{R \sigma}^{s}$ is obtained by exchanging $L
\leftrightarrow R$ in the r.h.s. of Eq.(\ref{ssl}). We observe
that $I_{R \sigma}^{s} \neq -I_{L \sigma}^{s}$ since the spin
current is not conserved in the presence of spin-flip
interactions. The second term on the r.h.s. of Eq.(\ref{itot})
describes the transient behavior and is expressed in terms of the
quantity
\begin{widetext}
\begin{eqnarray}
T_{\alpha}( \omega,t)&=& \sum_{\beta} U_{\beta}
\frac{e^{i(\omega+U_{\beta}-H_{QD}+\frac{i}{2}\Gamma)t}}{(\omega-H_{QD}+\frac{i}{2}\Gamma)(\omega+U_{\beta}-H_{QD}+\frac{i}{2}\Gamma)}
\Big{[} i\delta_{\alpha, \beta} - \Gamma_{\beta} \,
\frac{1}{\omega+U_{\beta}-H_{QD}+\frac{i}{2}\Gamma}
 \Big{]} +\hc \nonumber \\
&-&  \sum_{\beta} U_{\beta}^{2}
\frac{e^{-i(H_{QD}-\frac{i}{2}\Gamma)t}}{(\omega-H_{QD}+\frac{i}{2}\Gamma)(\omega+U_{\beta}-H_{QD}+\frac{i}{2}\Gamma)}
\,\Gamma_{\beta} \,
\frac{e^{i(H_{QD}+\frac{i}{2}\Gamma)t}}{(\omega-H_{QD}-\frac{i}{2}\Gamma)(\omega+U_{\beta}-H_{QD}-\frac{i}{2}\Gamma)}
\, .
\label{trans}
\end{eqnarray}
\end{widetext}
Few remarks about Eq.(\ref{trans}) are in order. In linear
response theory only the contribution of the first line remains.
Such contribution is responsible for transient oscillations which
are exponentially damped over a timescale $|\tau_{j}|$ and have
frequencies $\omega_{\alpha j} \sim |E_{F} +U_{\alpha} - h_{j}|$,
where $h_{j} +i \tau_{j}^{-1}$, $j=1,2$, are the two eigenvalues
of $H_{QD}+i\Gamma/2$. These oscillations originate from virtual
transitions between the resonant levels of the QD and the Fermi
level of the biased continua (\textit{resonant-continuum}). No
information about the intra-dot transitions
(\textit{resonant-resonant}) is here contained. Resonant-resonant
transitions are instead described by the contribution of the
second line of Eq.(\ref{trans}) which yields oscillations of
frequency $\omega_{1,2} = |h_{1}-h_{2}|$ damped as $\exp
(-t/\tau_{1,2})$ with $\tau_{1,2}^{-1} =
|\tau_{1}|^{-1}+|\tau_{2}|^{-1}$. Finally, it is straightforward
to verify that for $\Gamma_{\alpha}$ and $H_{QD}$ both diagonal in
spin space, Eq.(\ref{itot}) decouples into two identical formulae
(but with different parameters), which exactly reproduce the
well-known spinless result.\cite{jahuo,stefanucci}

Below we consider diagonal matrices $\Gamma_{\alpha \nu}$, in
agreement with the discussion at the end of Section \ref{sec1}.
Thus, all spin-flip scatterings occur in the quantum dot. The
ferromagnetic nature of the leads is accounted for by modeling the
$\Gamma$ matrices as\cite{rud}
\begin{eqnarray}
\mathrm{P} && \quad \left\{
\begin{array}{ccc}
  \Big{(}\Gamma_{\alpha \uparrow} \Big{)}_{\uparrow \uparrow} &=&
 \Gamma_{0 \alpha}(1+p) \quad \quad \alpha=L,R \\
  \Big{(}\Gamma_{\alpha \downarrow} \Big{)}_{\downarrow \downarrow} &=&
\Gamma_{0 \alpha}(1-p) \quad \quad \alpha=L,R \\
\end{array}
 \right. \nonumber \\
\mathrm{AP} && \quad \left\{
\begin{array}{ccc}
  \Big{(}\Gamma_{L \uparrow} \Big{)}_{\uparrow \uparrow} &=&
\Gamma_{0 L}(1+p)  \\
  \Big{(}\Gamma_{L \downarrow} \Big{)}_{\downarrow \downarrow} &=&
\Gamma_{0 L}(1-p)  \\
\Big{(}\Gamma_{R \uparrow} \Big{)}_{\uparrow \uparrow} &=&
\Gamma_{0 R}(1-p)  \\
  \Big{(}\Gamma_{R \downarrow} \Big{)}_{\downarrow \downarrow} &=&
\Gamma_{0 R}(1+p)  \\
\end{array}
 \right.
 \label{gammawbl}
 \end{eqnarray}
where all the remaining matrix elements are vanishing and $0<p<1$
is proportional to the polarization of the leads. Moreover we
assume that $\Gamma_{0 \alpha}  \equiv \Gamma_{0}$ does not depend
on $\alpha$, yielding a Left/Right symmetry in absence of
ferromagnetism.

\subsection{Normal case: quantum beats}
\label{sec3a}

In this Section we consider a quantum dot with an intra-dot energy
splitting $E_{z}$ between the $\uparrow$ and $\downarrow$ levels,
and an intra-dot spin-flip energy $V_{sf}$. The QD is coupled to
$L$ and $R$ normal electrodes, $p=0$. The spin splitting $E_{z}$
leads to two different transient frequencies and produces coherent
quantum beats in  both the total and spin currents.\cite{souza}
The effect of spin-flip scatterings on the coherent oscillations
is studied.

\begin{figure}[htbp]
    \vspace{0.2cm}
\includegraphics*[width=.47\textwidth,height=5.5cm]{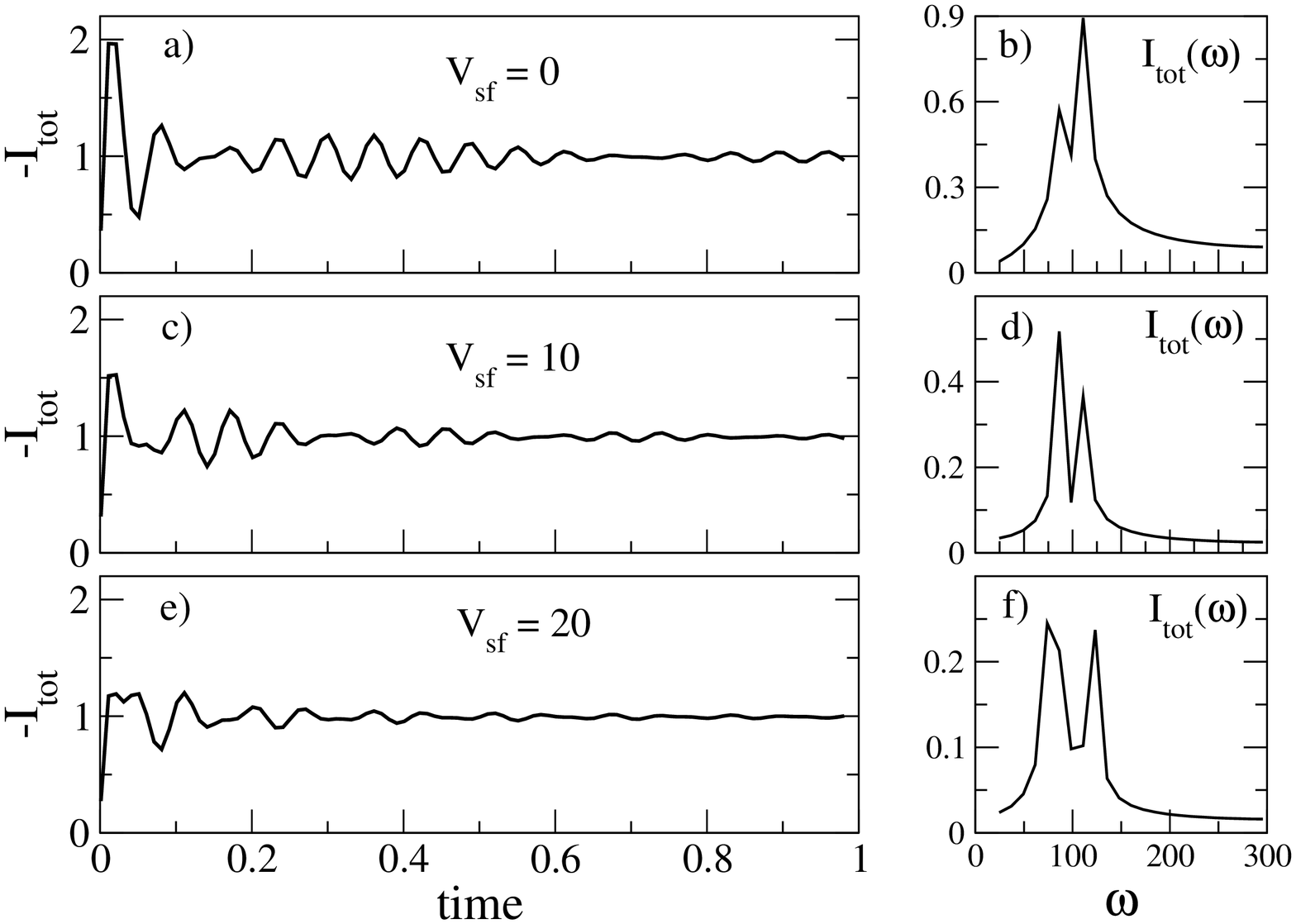}
\caption{$I_{tot}(t)$ and $I_{tot}(\omega)$ for $V_{sf}=0$ [panels
a)-b)], $V_{sf}=10$ [panels c)-d)], $V_{sf}=20$ [panels e)-f)].
The discrete Fourier transform is calculated using 50 equidistant
points of $I_{tot}(t)-I_{tot}(\infty)$ with $t$ in the range
(0.5,1). The rest of the parameters are $U_{L}=0$, $U_{R}=200$,
$\varepsilon_{d \sigma}=\theta (t) U_{R}/2 + \sigma E_{z}/2$ with
$E_{z}=10$, $\Gamma_{0}=1$ and inverse temperature $\beta = 100$.
} \label{wblbeats}
\end{figure}

\begin{figure}[htbp]
    \vspace{0.2cm}
\includegraphics*[width=.47\textwidth,height=5.5cm]{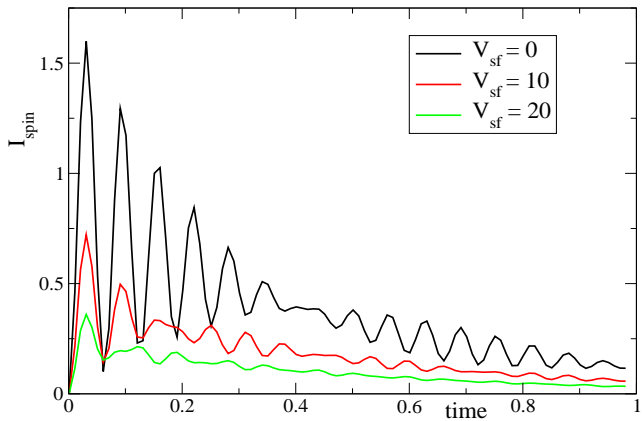}
\caption{(Colour online) $I_{spin}(t)$ for $V_{sf}=0, \,10, \,20$
and the same parameters as in Fig.\ref{wblbeats}.}
\label{wblbeatsb}
\end{figure}

To make contact with Ref.\onlinecite{souza}, we consider the same
parameters, i.e. $U_{L}=0$, $U_{R}=200$, $\varepsilon_{d
\sigma}=\theta (t) U_{R}/2 + \sigma E_{z}/2$ with $E_{z}=10$,
$\Gamma_{0}=1$ and inverse temperature $\beta = 100$. However the
intra-dot spin-flip coupling $V_{sf}$ is, in our case, non zero.
In Fig.\ref{wblbeats} we show the time-dependent current
$I_{tot}(t)$ and the discrete Fourier transform,\cite{stefanucci2}
$I_{tot}(\omega)$, of $I_{tot}(t)-I_{tot}(\infty)$ for different
values of $V_{sf}= 0,\,10,\,20$. The spin current $I_{spin}(t)$ is
displayed in Fig.\ref{wblbeatsb}. The small tunneling rate
$\Gamma$ leads to a very long transient regime, a property which
allows us to observe well defined structures in the Fourier
spectrum of the current. At $V_{sf}=0$ the frequencies of both
$I_{\ua}(t)$ and $I_{\da}(t)$ are $\omega_{\alpha
1}=\omega_{\alpha 2} = |U_{R}/2 \pm E_{z}/2|= 95,105$, in
agreement with the results of Ref.\onlinecite{souza}. This is
confirmed by the Fourier analysis of $I_{tot}$ which is displayed
in panel b) of Fig.\ref{wblbeats}. As $V_{sf}$ is increased the
frequencies of the transient oscillations change according to
$|U_{R}/2 \pm \sqrt{E_{z}^{2}/4+V_{sf}^{2}}|$, see panels d) and
f) of Fig.\ref{wblbeats}, and the amplitude of the quantum beats
in $I_{tot}$ and $I_{spin}$ is suppressed. Such suppression is due
to the fact that the Hamiltonian of the whole system is
spin-diagonal if we choose the quantization axis along
\begin{equation}
\hat{\xi} =\hat{z}\cos\th+\hat{x}\sin\th\, ,
\label{qa}
\end{equation}
with $\cos\th=E_z/\sqrt{E_{z}^{2}+4 V_{sf}^{2}}$. This follows
from the invariance of the normal electrode Hamiltonians ($p=0$)
and of the tunneling Hamiltonian ($\Gamma_{\alpha \sigma} =
\Gamma_{0} \bfone_{2 \times 2}$) under rotations in spin-space.
Therefore, the time-dependent spin current measured along
$\hat{z}$ is smaller the larger $V_{sf}$ is. Of course the spin
current measured along the quantization axis $\hat{\xi}$ is not
suppressed by changing $V_{sf}$.

Despite the quantum beats phenomenon and its suppression with
increasing $V_{sf}$ are well captured within the WBLA, only a
sub-set of transient frequencies can be predicted in this
approximation. To describe transitions between the top/bottom of
the electrode bands and the resonant levels requires a more
realistic treatment of the leads Hamiltonian, see Section
\ref{sec3}.

\subsection{Ferromagnetic case: TMR}

The finite spin-polarization $p$ of the electrodes breaks the
invariance of $H_{L}$ and $H_{R}$ under rotations in spin-space.
Thus the one-particle states become true spinors for nonvanishing
$V_{sf}$. Using Eq.(\ref{ssl}) we calculate the steady-state total
currents both in the P and AP configurations, which we denote as
$I_{P}$ and $I_{AP}$, and study the TMR,
\begin{equation}
\mathrm{TMR}=\frac{I_{P}-I_{AP}}{I_{AP}} \, .
\label{tmr}
\end{equation}
For $V_{sf}=0$ the TMR is always positive and its sign can be
inverted only during the transient regime at sufficiently low
temperatures.\cite{souza} The  positiveness of the steady state
TMR is due to the fact that $\Gamma_{0 L}=\Gamma_{0 R}$. It is
possible to show that for $\Gamma_{0L} \gg \Gamma_{0R} $ (or
$\Gamma_{0L} \ll \Gamma_{0R} $) and $\varepsilon_{d \sigma}=0$,
the $\mathrm{TMR} \approx -p^{2}$ in linear response theory
yielding the so-called \textit{resonant inversion} of the
TMR.\cite{tsymbal}

\begin{figure}[h!]
\includegraphics*[width=.47\textwidth]{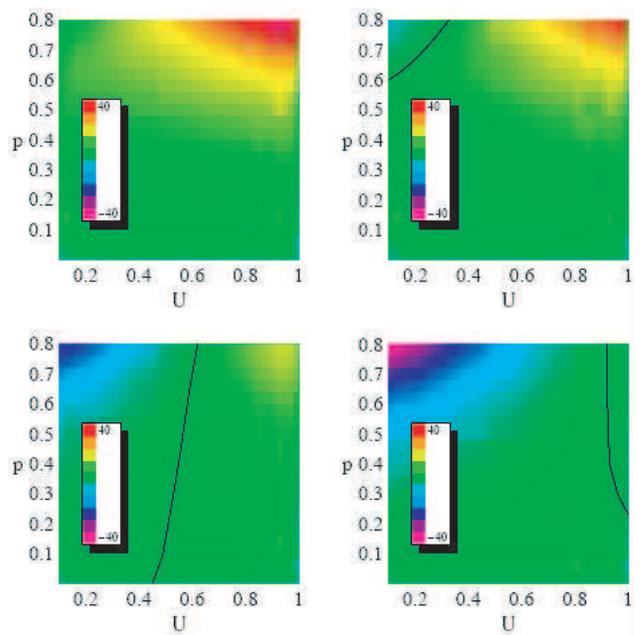}
\caption{(Colour online) Contour plot of the TMR at the
steady-state in units of $10^{-2}$ as a function of $U$ and $p$
for different values of $V_{sf}=0.3$ (top-left), 0.33 (top-right),
0.4 (bottom-left), 0.47 (bottom right). The boundary TMR $=0$ is
displayed with a black line. The remaining parameters are
$\Gamma_{0}=0.5$, $\varepsilon_{d \sigma}=0$, $\beta=100$. }
\label{two}
\end{figure}

\begin{figure}[htbp]
    \vspace{0.2cm}
\includegraphics*[width=.2\textwidth,height=.2\textwidth]{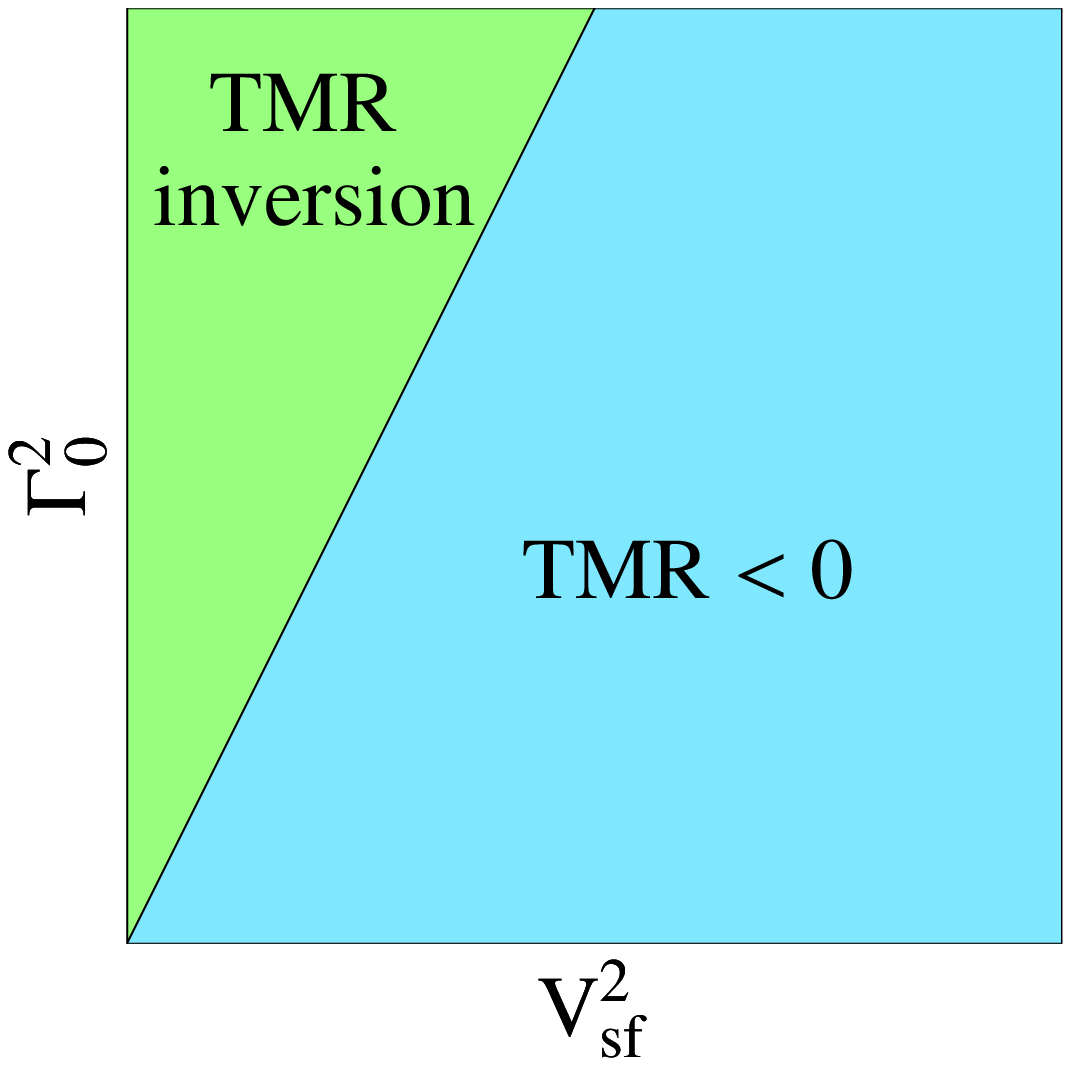}
\hspace{0.5cm}
\includegraphics*[width=.23\textwidth,height=.2\textwidth]{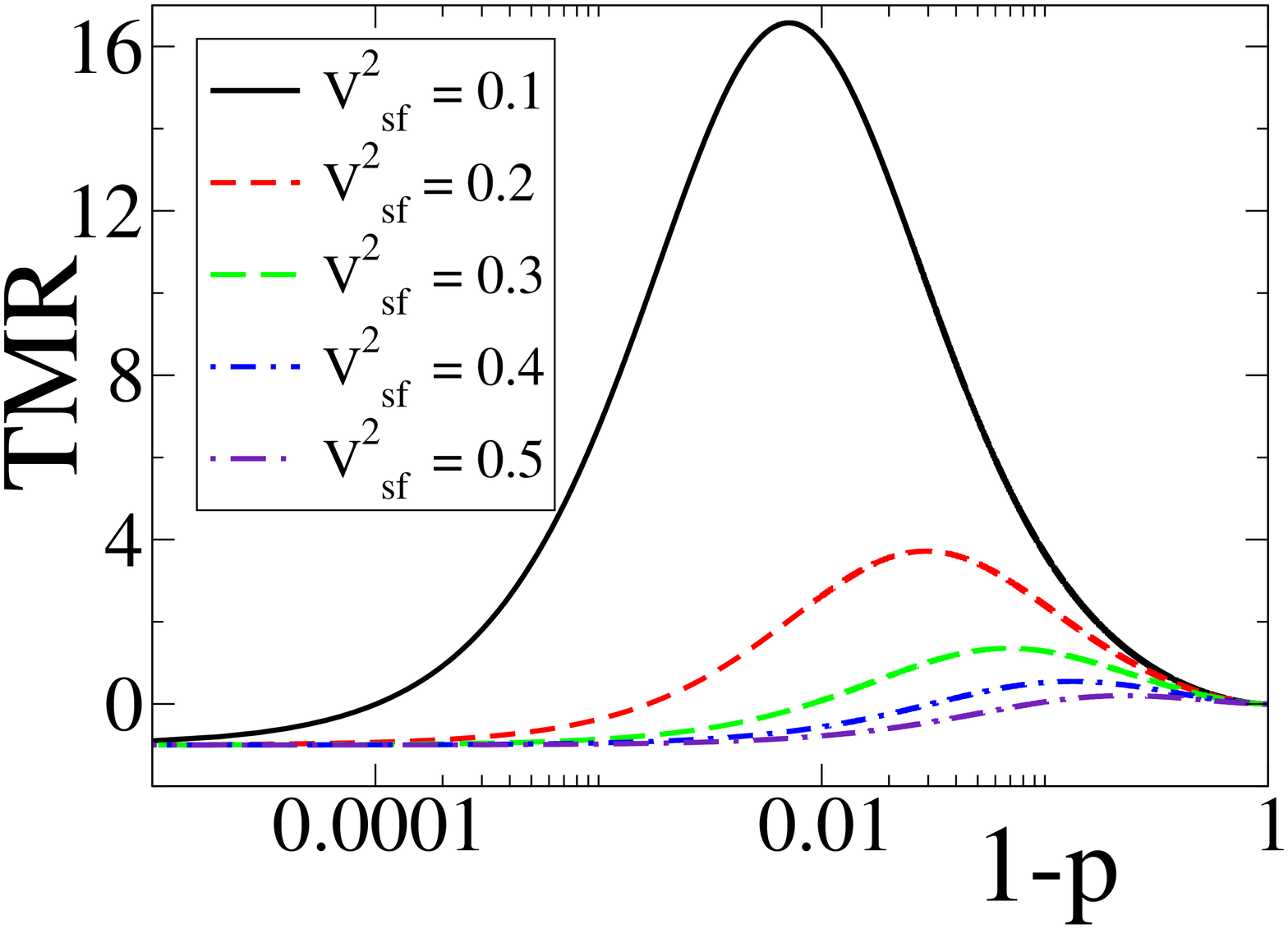}
\caption{(Colour online) Left panel: in the region $\Gamma_{0}^{2}
< 2 V_{sf}^{2}$ the TMR is always negative while in the region
$\Gamma_{0}^{2} > 2 V_{sf}^{2}$ the TMR changes sign depending on
the value of the polarization $p$. Right panel: TMR as a function
of $p$ for $\Gamma_{0}=1$ and different values of $V_{sf}$. }
\label{tmrwbllr}
\end{figure}

In Fig.\ref{two} we display the contour plot of the TMR in the
parameter space spanned by the bias voltage $U=U_{L}=-U_{R}$ and
the polarization $p$ for different values of the spin-flip energy
$V_{sf}$ and for $E_{z}=0$, $\Gamma_{0}=0.5$ and $\beta =100$. We
observe that by increasing $V_{sf}$, \textit{a region of negative}
TMR \textit{appears }and becomes wider the larger $V_{sf}$ is. The
region in which the TMR is appreciably different from zero
($|\mathrm{TMR}| \gtrsim 0.05$ ) is confined to the high
magnetization regime, i.e. $p \gtrsim 0.4$. The minimum of the TMR
is reached for the largest value of $V_{sf}$ at small $U$ and
large $p$, see panel d) of Fig.\ref{two}. For small biases the
sign inversion of the TMR can be understood by calculating the
currents $I_{P}$ and $I_{AP}$ in linear response. By expanding
Eq.(\ref{ssl}) to first order in $U$ and using the $\Gamma$
matrices of Eq.(\ref{gammawbl}) one can show that to leading order
in the bias
\begin{equation}
\mathrm{TMR}=p^{2}\frac{(1-p^{2})\Gamma_{0}^{4}-(1-p^{2})\Gamma_{0}^{2}V_{sf}^{2}-2V_{sf}^{4}}
{[V_{sf}^{2}+(1-p^{2})\Gamma_{0}^{2}][(1+p^{2})V_{sf}^{2}+(1-p^{2})\Gamma_{0}^{2}]}
\, .
\end{equation}
For any given $V_{sf}$ and $\Gamma_{0}$ the denominator in the
above expression is positive while the numerator can change sign.
The regions of posite/negative TMR are displayed in the left side
of Fig.\ref{tmrwbllr}. For $\Gamma_{0}^{2}<2V_{sf}^2$ the TMR is
negative independent of the polarization $p$. This follows from
the fact that the tunneling time is larger than the time for an
electron to flip its spin, thus favoring the AP alignment. Along
the boundary $\Gamma_{0}^{2}=2V_{sf}^2$ the TMR is zero only for
$p=0$ and negative otherwise. In the region
$\Gamma_{0}^{2}>2V_{sf}^2$ there exists a critical value of the
polarization,
\begin{equation}
p_{c}=\sqrt{1-\frac{2
V_{sf}^{4}}{\Gamma_{0}^{4}-\Gamma_{0}^{2}V_{sf}^{2}}} \, ,
\end{equation}
such that the TMR is positive for $p>p_{c}$ and negative for
$p<p_{c}$. It is worth to emphasize that at the boundary
$V_{sf}=0$ the TMR is positive for any finite value of $p$ and
zero for $p=0$. This analysis clearly shows that the sign of the
TMR can be reversed without resorting to asymmetric couplings
$\Gamma_{L} \neq \Gamma_{R}$, provided spin-flip processes are
included. The right side of Fig.\ref{tmrwbllr} displays the TMR as
a function of the polarization for different values of the ratio
$V_{sf}^{2}/\Gamma_{0}^{2} < 1/2$ (region of TMR inversion).

We also have investigated the temperature dependence of the TMR.
This is relevant in the light of practical applications, where a
large TMR is desirable at room temperature. The increase of
temperature tends to suppress the negative values of the TMR while
the positive values remain almost unaffected. When $\beta \approx
1$ the region of the TMR inversion disappears. The positiveness of
the TMR at high temperatures has been already observed in
Ref.\onlinecite{rud}.

In the next Section we provide an exact treatment of the 1D leads
and illustrate differences and similarities with the results
obtained so far.

\section{Results for one-dimensional leads}
\label{sec3}

The numerical results contained in this Section are obtained by
computing the exact time-evolution of the system in
Eq.(\ref{htot}) with a finite number $N$ of sites in both $L$ and
$R$ leads. \cite{finiteleads1,finiteleads2,finiteleads3} Let us
define the biased Hamiltonian at positive times as
\begin{equation}
H_{bias}(t)=H_{0}+H(t)
\end{equation}
with
\begin{equation}
H(t)= \theta (t) \sum_{\sigma =\uparrow,\downarrow}\sum_{j=1}^{N}
U_{\alpha}(t) c^{\dagger}_{\alpha j \sigma} c_{\alpha j \sigma} \,
.
\end{equation}
Both $H_{0}$ and $H_{bias}$ have dimension $2(2N+1)$, where the
factor of 2 accounts for the spin.

\begin{figure}[htbp]
    \vspace{0.2cm}
\includegraphics*[width=.4\textwidth,height=4.5cm]{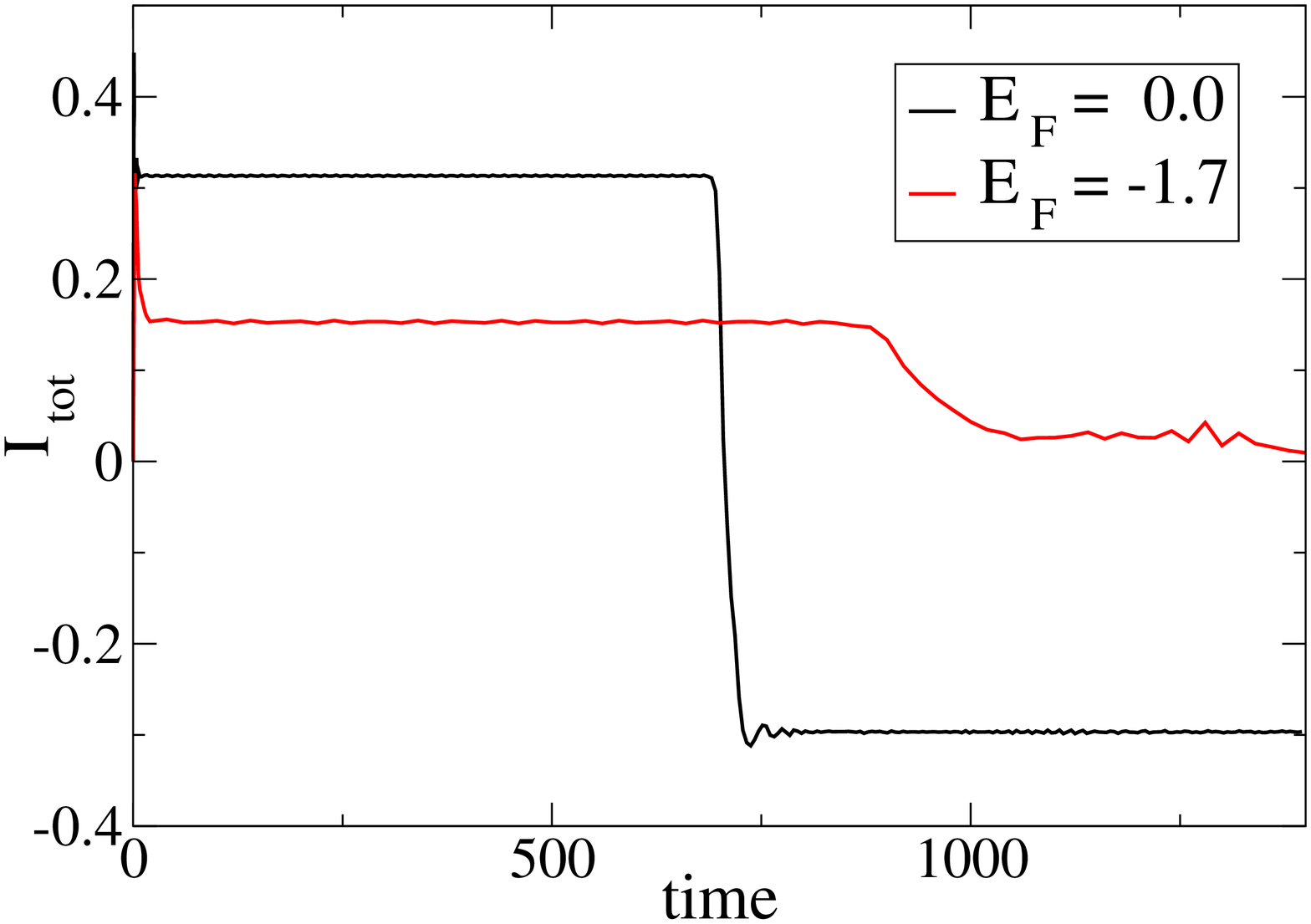}
\caption{(Colour online) $I_{\uparrow}(t)$ at zero temperature for
$N=700$, $U_{L}=0.5$, $U_{R}=0$, $E_{F}=0$ ($T_{max} \approx
700$), and $E_{F}=-1.7$ ($T_{max} \approx 870$). The rest of the
parameters are $\varepsilon_{d \sigma}=0$, $V=-1$, $V_{sf}=0$. The
current reaches a well-defined steady state before the occurrence
of finite-size effect ($t \gtrsim T_{max}$).} \label{finite}
\end{figure}

We use the partition-free approach\cite{cini} and specialize to a
sudden switching of a constant bias. Accordingly, we first
calculate the equilibrium configuration by numerically
diagonalizing $H_{0}$, and then we evolve the lesser Green's
function as
\begin{equation}
G^{<}(t,t')=ie^{-iH_{bias}t}f(H_{0})e^{iH_{bias}t'} \, ,
\label{gless}
\end{equation}
with $f$ the Fermi distribution function. The spin-polarized
current flowing across the bond $m-n$ is then calculated from
\begin{equation}
I_{m,n,\sigma}(t) =2 \sum_{\sigma'} [H_{0}]_{m\sigma;n\sigma'} \,
\Im\{-i[G^{<}(t,t)]_{n\sigma';m\sigma} \} \, ,
\end{equation}
where $[\dots]_{m\sigma;n\sigma'}$ denotes the matrix element
associated to site $m$ with spin $\sigma$ and site $n$ with spin
$\sigma'$, while $\Im$ stands for the imaginary part. The above
approach allows us to reproduce the time evolution of the
infinite-leads system provided one evolves up to a time $T_{max}
\approx 2N/v$, where $v$ is the maximum velocity for an electron
with energy within the bias window. Indeed for $t \gtrsim T_{max}$
high-velocity electrons have time to propagate till the far
boundary of the leads and back yielding undesired finite-size
effects in the calculated current, see Fig.\ref{finite}. For this
reason we set $N$ much larger than the time at which the steady
state (or stationary oscillatory state in the case of AC bias) is
reached. We tested this method by comparing our numerical results
against the ones obtained in
Refs.\onlinecite{stefanucci3,stefanucci2} where the leads are
virtually infinite, and an excellent agreement was found for $t
\lesssim T_{max}$. Moreover, the value of the current at the
steady-state agrees with the Landauer formula with high numerical
accuracy.

Below we study the spin-polarized current at the left interface
$I_{\sigma} = I_{1L,d,\sigma} $ flowing between the first site of
the lead $L$ ($m=1L$) and the QD ($n=d$) with spin $\sigma$. We
stress that our approach is not limited to the calculation of the
current through a specific bond, as we have access to the full
lesser Green's function. Sensible electron densities and currents
in the vicinity of the QD can be extracted and their calculation
requires the same computational effort.

\subsection{Normal case: quantum beats}
\label{sec4}

We study the quantum beats phenomenon for normal 1D leads ($\Delta
\varepsilon =0$) and intra-dot spin-splitting $E_{z}$ for
different values of the spin relaxation energy $V_{sf}$. The
comparison between the results obtained within the WBLA and with
1D tight-binding leads will turn out to be very useful to
elucidate advantages and limitations of the former approach.

\begin{figure}[htbp]
    \vspace{0.2cm}
\includegraphics*[width=.47\textwidth,height=5.5cm]{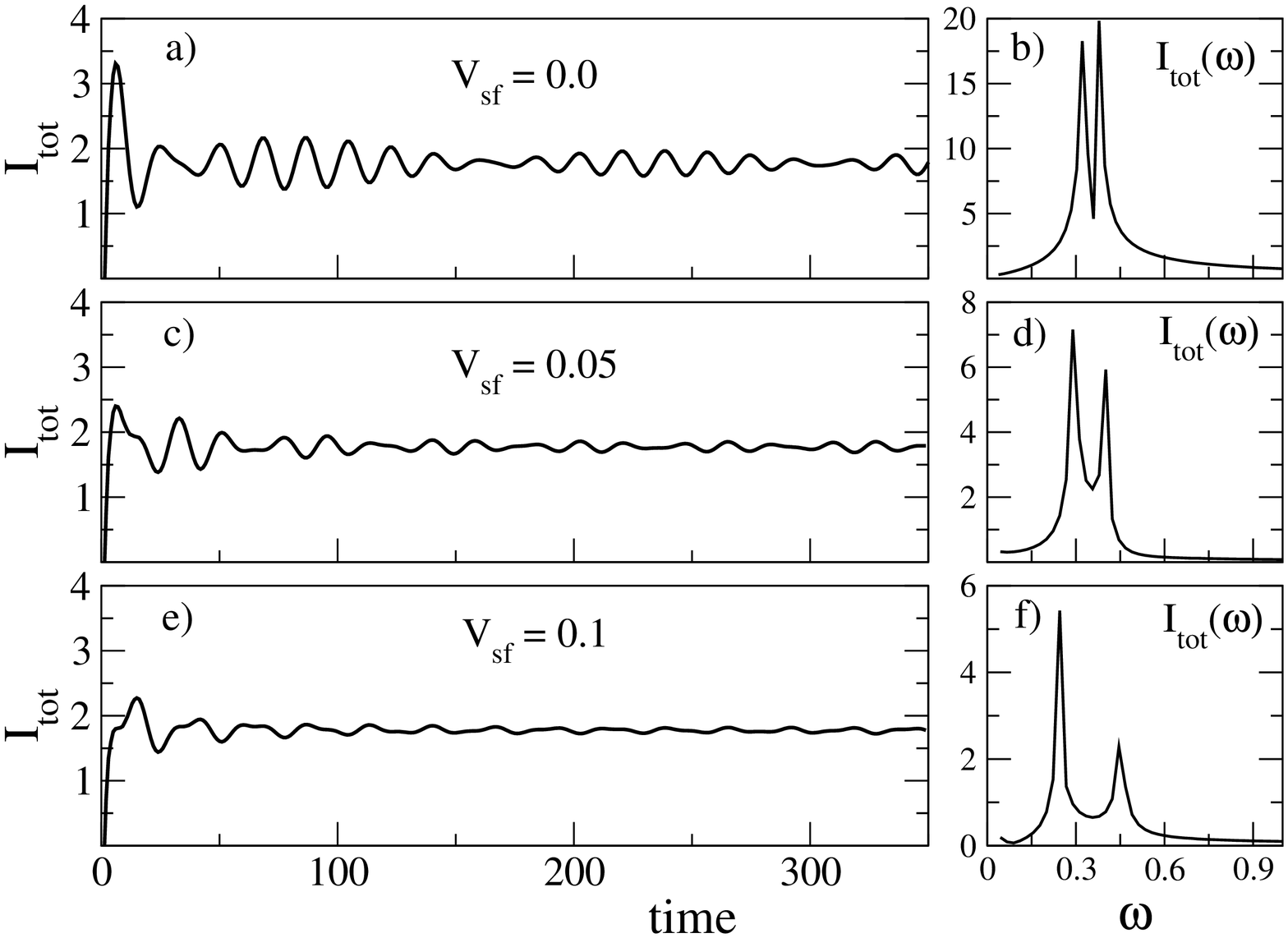}
\caption{(Colour online) $I_{tot}(t)$ and $I_{tot}(\omega)$ for
 $V_{sf}=0$ [panels a)-b)], 0.05 [panels c)-d)], 0.1 [panels e)-f)]
obtained with a bias  $U_{L}=0.7$, $U_{R}=0$ applied to leads of
length $N=450$. The discrete Fourier transform is calculated using
280 equidistant points of $I_{tot}(t)-I_{tot}(\infty)$ with $t$ in
the range (70,350). The rest of the parameters are $V=0.03$,
$\varepsilon_{d \sigma}=\theta (t) U_{L}/2 + \sigma E_{z}/2$,
$E_{z}=0.04$ and zero temperature.} \label{beats}
\end{figure}

\begin{figure}[htbp]
    \vspace{0.2cm}
\includegraphics*[width=.47\textwidth,height=5.5cm]{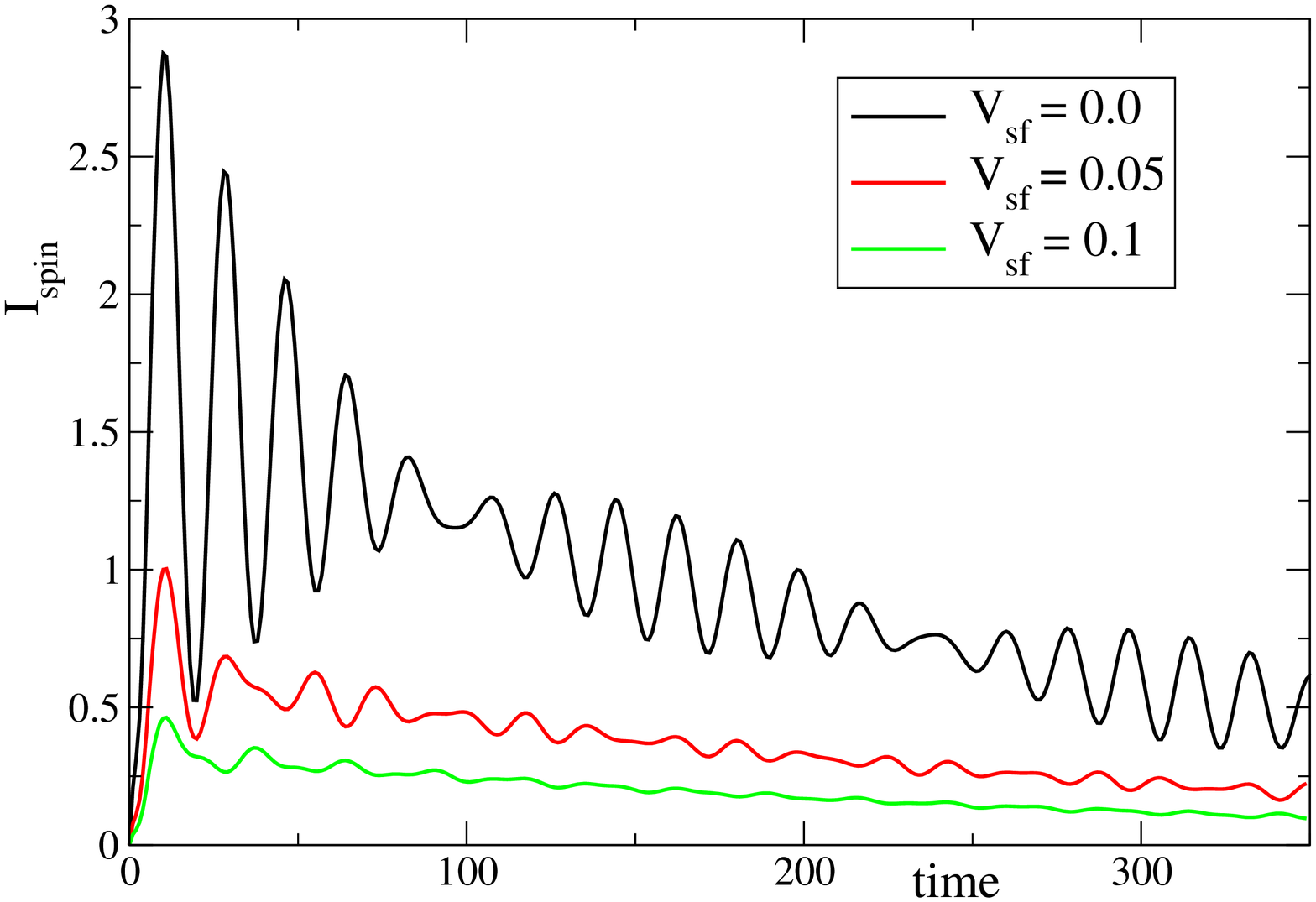}
\caption{(Colour online) $I_{spin}(t)$ for the same system
parameters of Fig.\ref{beats}. } \label{beats1}
\end{figure}

In Fig.\ref{beats} we show the time-dependent currents
$I_{tot}(t)$ and its discrete Fourier transform $I_{tot}(\omega)$
at zero temperature using the parameters $U_{L}=0.7$, $U_{R}=0$,
$V=0.03$, $\varepsilon_{d \sigma}=\theta (t) U_{L}/2 + \sigma
E_{z}/2$, with $E_{z}=0.04$, for different values of $V_{sf}=0, \,
0.05, \, 0.1$. The spin-current for the same parameters is
displayed in Fig.\ref{beats1}. The value of the dot-lead hopping
$V$ corresponds to an effective (energy-dependent) tunneling rate
$\Gamma$ of the order of $10^{-3}$. Thus both the bias and the
energy spin-splitting are of the same order of magnitude as the
ones used in Section \ref{sec3} if measured in units of $\Gamma$.
As expected the current reaches a steady-state in the long time
limit, since the bias is constant and $H_{bias}$ does not have
bound states.\cite{finiteleads2,stefbound} However, the weak links
between the QD and the leads allows us to study a
pseudo-stationary regime, in which the current displays
well-defined quantum beats.

At $V_{sf}=0$, the dominant frequency of the spin-polarized
current $I_{\ua}$ is $\omega^{0}_{\uparrow}=|U_{L}/2 + E_{z}/2|
\approx 0.37$ while $I_{\downarrow}$ oscillates with a dominant
frequency $\omega^{0}_{\da}=|U_{L}/2 - E_{z}/2| \approx 0.33$, see
panel b) of Fig.\ref{beats}. As in the WBLA, the difference
between $\omega_{\ua}$ and $\omega_{\da}$ leads to quantum beats
in both $I_{tot}$ and $I_{spin}$ [see panel a) of Fig.\ref{beats}
and Fig.\ref{beats1}]. For nonvanishing $V_{sf}$ the two
fundamental frequencies $\omega^{0}_{\sigma}$ renormalizes as
$\omega^{0}_{\sigma} \ra \omega_{\sigma} =|U_{L}/2 + \sigma
\sqrt{E_{z}^{2}/4+V_{sf}^{2}}|$. The system Hamiltonian is no
longer diagonal along the quantization axis $\hat{z}$ and
$I_{\sigma}$ acquires the second frequency $\omega_{-\sigma}$
besides the original (but renormalized) one $\omega_{\sigma}$. We
also observe that as $V_{sf}$ increases, the amplitude of the
quantum beats in $I_{tot}$ and $I_{spin}$ is suppressed, similarly
to what happen by treating the leads in the WBLA.

\begin{figure}[htbp]
    \vspace{0.2cm}
\includegraphics*[width=.47\textwidth,height=5.5cm]{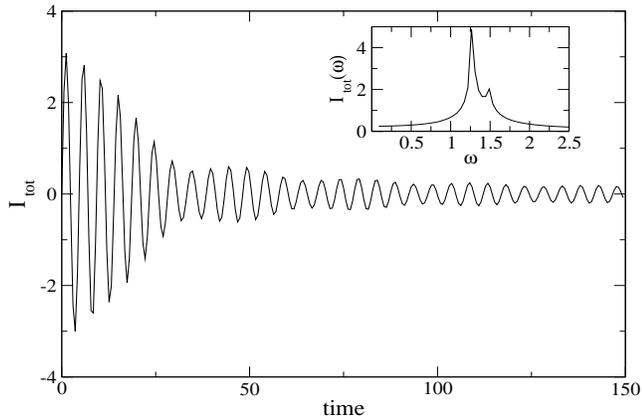}
\caption{$I_{tot}(t)$ in units of $10^{-6}$ for $U_{R}=0.4$,
$U_{L}=0$ applied to leads of length $N=250$. The inset shows
$I_{tot}(\omega)$ in units of $10^{-5}$ calculated using 138
equidistant points of $I_{tot}(t)-I_{tot}(\infty)$ with $t$ in the
range (12,150). The rest of the parameters are $E_{z}=0$,
$E_{F}=-1.8$, $\varepsilon_{d}=-0.5$, $V=0.02$, $V_{sf}=0$ and
zero temperature. } \label{beats1D}
\end{figure}

We would like to end this Section by pointing out that, for leads
with a finite bandwidth the current might display extra
oscillation frequencies corresponding to transitions either from
or to the top/bottom of the bands, an effect which cannot be
captured within the WBLA. We have investigated this scenario by
changing the input parameters in such a way that the transitions
from the resonant level of the QD to the bottom of the band are
energetically favored. We set $U_{R}=0.4$, $U_{L}=0$, $E_{z}=0$,
$E_{F}=-1.8$, $\varepsilon_{d}=-0.5$, $V=0.02$ and $V_{sf}=0$. The
corresponding (spin-independent) current and its Fourier spectrum
are shown in Fig.\ref{beats1D}. One can clearly see two well
defined peaks at energies  $\omega_{1} \approx 1.3$ and
$\omega_{2} \approx 1.5$. As expected, one of these frequencies
corresponds to a transition from the resonant level to the Fermi
energy of the $L$ lead, i.e. $|E_{F}-\varepsilon_{d}|=1.3$.
Transitions between the resonant level and the Fermi energy of the
$R$ lead, i.e. $|E_{F}+U_{R}-\varepsilon_{d}|=0.9$, are strongly
suppressed since the current is measured at the $L$ interface.
Thus the second peak has to be ascribed to a transition which
involves the bottom of the band $E_{B}=-2$, specifically the
transition of energy $|E_{B}-\varepsilon_{d}|=1.5$. This kind of
features in the Fourier spectra of the transients points out the
limitations of the WBLA and might be experimentally observed in
QD's connected to narrow-band electrodes.

\subsection{Normal case: engineering the spin-polarization}
\label{ivb}

In this Section we exploit the full knowledge of the
time-dependent response of the system in order to engineer the
spin-polarization of the total current. In particular we are
interested in maintaining the polarization ratio
\begin{equation}
r(t)=2\frac{I_{tot}(t) I_{spin}(t)}{
I_{tot}^{2}(t)+I_{spin}^{2}(t)}
\end{equation}
above some given value in a sequence of time windows of desired
duration. The quantity $r$ is $\pm 1$ for fully polarized currents
and zero for pure charge or spin currents.
\begin{figure}[htbp]
    \vspace{0.2cm}
\includegraphics*[width=.47\textwidth,height=5.5cm]{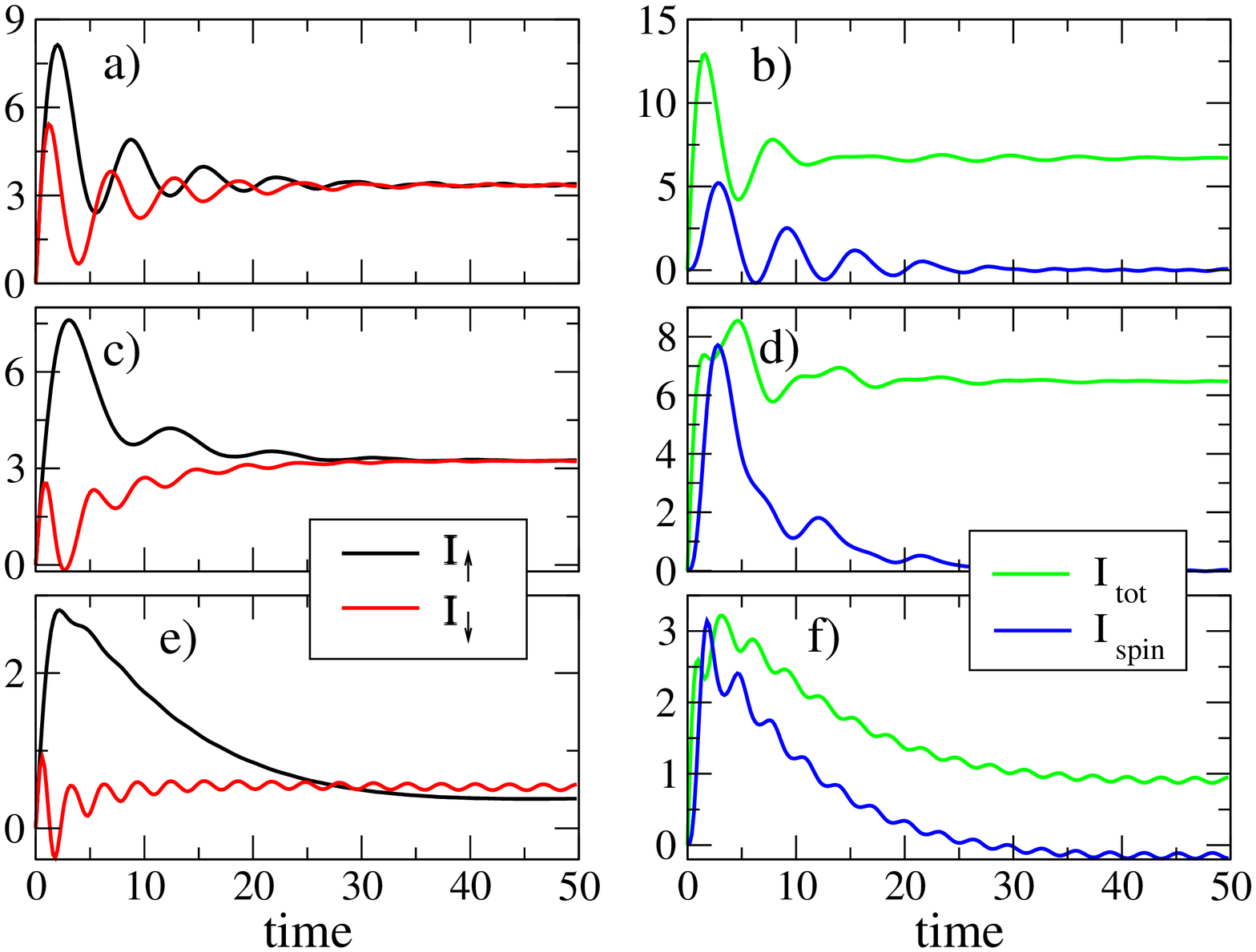}
\caption{(Colour online) Currents in units of $10^{-2}$ for
$U_{L}=-U_{R}=1$ applied to leads of length $N=120$ for different
values of $E_{z}=0.1$ [panels a)-b)], 0.6 [panels c)-d)], 2
[panels e)-f)]. The rest of the parameters are $V=0.2$, $V_{sf}=0$
and zero temperature.} \label{engconstant}
\end{figure}

In Fig.\ref{engconstant} we show the time-dependent currents
$I_{\ua}(t)$ and $I_{\da}(t)$ as well as $I_{tot}(t)$ and
$I_{spin}(t)$ at zero temperature for different values of
$E_{z}=0.1, \, 0.6, \, 2$ and $U_{L}=-U_{R}=1$, $V=0.2$, and
$V_{sf}=0$. In the long-time limit the values of $r$ increases
steadily as $E_{z}$ is increased, and remains below $10^{-1}$ for
$E_{z} \lesssim 2$. The polarization ratio can, however, be much
larger in the transient regime and its maximum has a nontrivial
dependence on $E_{z}$. At finite values of $E_{z}$ there is an
initial unbalance of the spin up and down densities. This implies
that for small times $I_{\downarrow}$ is suppressed and delayed
with respect to $I_{\uparrow}$, since spin $\uparrow$ electrons
can freely flow while there cannot be any spin $\downarrow$ flow
before the initial spin $\downarrow$ electrons have left the
QD.\cite{souza} This is the so called Pauli blockade phenomenon
and is responsible for a recoil of $I_{\downarrow}$ during the
rise of $I_{\uparrow}$. In panel a)-b) of Fig.\ref{engconstant} we
consider the case $E_{z}=0.1$. Both $I_{\ua}$ and $I_{\da}$
overshoot their steady-state value and oscillate with rather sharp
maxima and minima during the initial transient, see panel a) of
Fig.\ref{engconstant}. However, the oscillation frequencies are
very similar  (small $E_{z}$) and the time at which $I_{\ua}$ has
a maximum in correspondence to a minimum of $I_{\da}$ occurs when
the amplitude of their oscillations is already considerably
damped. On the other hand, for $E_{z}=0.6$, see panel c)-d) of
Fig.\ref{engconstant}, there is a synergy between the Pauli
blockade phenomenon and the frequency mismatch. At small $t$ such
synergy generates large values of the ratio $r \approx 1$ despite
$r(t\ra\inf)$ is vanishingly small. By increasing further the
value of $E_z$ we observe a long overshoot of $I_{\ua}$ while
$I_{\da}$ oscillates with high frequency, see panel e)-f) of
Fig.\ref{engconstant} where $E_{z}=2$. The ratio $r(t)$ is a
smooth decreasing function of time for $t\gtrsim 3$ and approaches
the value of about $10^{-2}$ for $t\ra\inf$. This behavior differs
substantially from the one obtained for $E_{z}=0.6$ where $r(t)$
has a sharp peak for $0\lesssim t\lesssim 3$ and is very small
otherwise. Below we show how one can exploit this kind of
transient regimes to maintain persistently a large value of $r$.

We consider the optimal case $E_{z}=0.6$ and apply a pulsed bias
with period $T$ and amplitude $U_{L}=-U_{R}=1$ in lead $L$ and $R$
respectively. The period $T$ of $H_{bias}$ is tailored to maintain
the polarization ratio $r(t)$ above $\sim 0.5$ in a finite range
of the period. For time-dependent biases Eq.(\ref{gless}) has to
be generalized, as the evolution operator is no longer the
exponential of a matrix. We discretize the time and calculate the
lesser Green's function according to
\begin{equation}
G^{<}(t_{n},t_{n}) \approx   e^{-i H_{bias}(t_{n})\Delta t} \,
G^{<}(t_{n-1},t_{n-1}) \,  e^{i H_{bias}(t_{n})\Delta t} \, ,
\end{equation}
where $t_{n}=n\Delta t$, $\Delta t$ is the time step, $n$ is a
positive integer and $G^{<}(0,0) = i f(H_{0})$.

\begin{figure}[htbp]
    \vspace{0.2cm}
\includegraphics*[width=.47\textwidth,height=3.5cm]{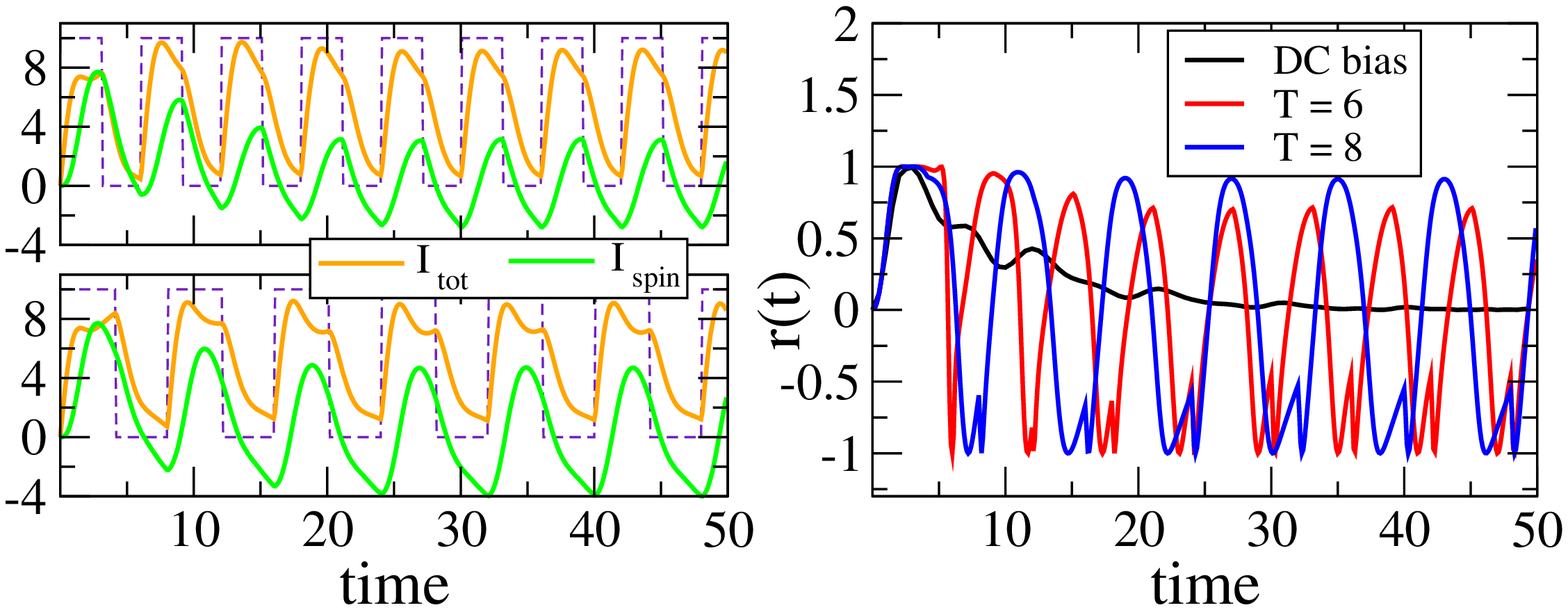}
\caption{(Colour online) $I_{tot}(t)$ and  $I_{spin}(t)$ in units
of $10^{-2}$ for a pulsed bias of period $T=6$ (top-left panel)
and $T=8$ (bottom-left panel) of amplitude $U_{L}=-U_{R}=1$
applied to leads of length $N=120$. The polarization ratio $r(t)$
is displayed in the right panel. The rest of the parameters are
the same as in Fig.\ref{engconstant}. The dotted lines in the left
panel denote the pulsed bias in the left lead in units of
$10^{-1}$. The numerical calculations have been performed with
$\Delta t =0.1$. } \label{engpulse}
\end{figure}

In Fig.\ref{engpulse} we plot two time-dependent responses for
$T=6$ and $T=8$. The pulsed bias produces an alternate $I_{tot}$
and $I_{spin}$ whose amplitude depends on $T$. We note that the
amplitude of $I_{tot}$ is of the same order of magnitude of the
steady state value $I_{tot}(t \ra \inf)$ attained in
Fig.\ref{engconstant} panel d) for constant bias. On the contrary
the amplitude of $I_{spin}$ is two orders of magnitude larger than
the corresponding steady-state value. The polarization $r$ cannot
be maintained as large as $r\approx 1$ (maximum value of $r$
during the transient) due to an unavoidable damping. However, the
value of $r$ is above 0.5 in a time window of 1.4 for $T=6$ (with
a maximum value $r\sim 0.75$) and $2.6$ for $T=8$ (with a maximum
value $r\sim 0.9$) in each period, see Fig.\ref{engpulse}.

\subsection{Ferromagnetic case: TMR}

\begin{figure}[htbp]
    \vspace{0.2cm}
\includegraphics*[width=.47\textwidth,height=5.5cm]{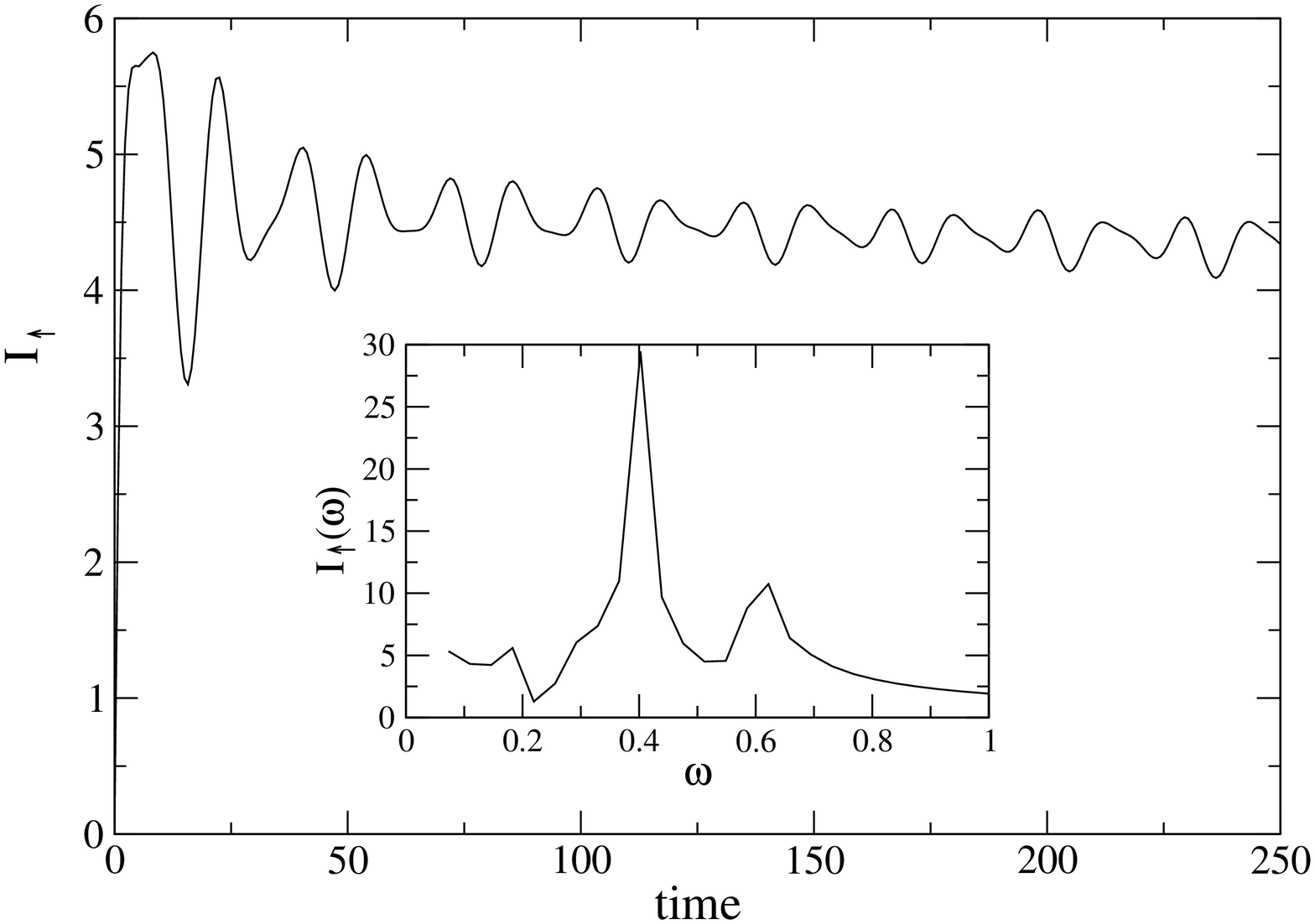}
\caption{$I_{\uparrow}(t)$ in units of $10^{-4}$ for $U_{L}=1$,
$U_{R}=0$ applied to leads of length $N=350$. The inset shows the
discrete Fourier transform $I_{\ua}(\w)$ calculated using 265
equidistant points of $I_{\ua}(t)-I_{\ua}(\infty)$ with $t$ in the
range (35,300). The rest of the parameters are
$\varepsilon_{d\sigma}= \theta(t) U_{L}/2 +\sigma E_{z}/2$,
$E_{z}=0.04$, $V=0.02$, $V_{sf}=0.1$ and zero temperature.
 } \label{ferrobeats}
\end{figure}

The spin-dependent band-structure of the leads introduces another
frequency dependence in the currents. Both $I_{\uparrow}$ and
$I_{\downarrow}$ display a coherent oscillation of frequency
$\omega_{1,2}=|h_{1}-h_{2}|=2 \sqrt{E_{z}^{2}/4+V_{sf}^{2}}$,
where $h_{1,2}$ are the eigenvalues of the isolated QD. Such
behavior cannot be observed in the normal case, as the model is
diagonal along the quantization axis $\hat{\xi}$ of Eq.(\ref{qa})
and no transitions between states of opposite polarization along
$\hat{\xi}$ can occur. In Fig.\ref{ferrobeats} we display the
spin-up current at zero temperature in the P configuration for
$\Delta \varepsilon =1.5$, $U_{L}=1$, $U_{R}=0$,
$\varepsilon_{d\sigma}= \theta(t) U_{L}/2 +\sigma E_{z}/2$,
$E_{z}=0.04$, $V=0.02$, and $V_{sf}=0.1$. Besides the frequencies
$\omega_{L1}=|E_{F}+U_{L}-h_{1}|\approx0.6$ and
$\omega_{L2}=|E_{F}+U_{L}-h_{2}|\approx0.4$ already discussed in
Section \ref{sec3a}, one can see the appearance of the new
frequency $\omega_{1,2}\approx0.2$.

Next, we study the steady-state regime in both the P and AP
configurations and calculate the TMR for different values of the
bias voltage $U$, the band spin splitting $\Delta \varepsilon$ and
the spin-flip energy $V_{sf}$. Analogies and differences with the
case of wide-band leads will be discussed.

\begin{figure}[h!]
\includegraphics*[width=.47\textwidth]{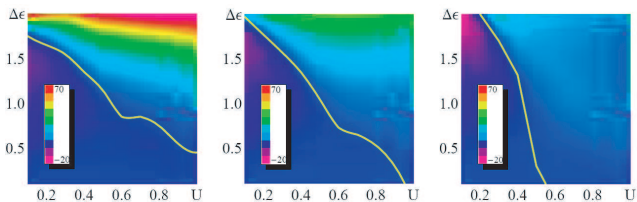}
\caption{(Colour online) Contour plot of the TMR at the
steady-state in units of $10^{-2}$ as a function of $U$ and
$\Delta \varepsilon$ for different values of $V_{sf}=0$ (left),
0.25 (center), 0.5 (right). The boundary TMR $=0$ is displayed
with a white line. The remaining parameters are $V=0.5$,
$\varepsilon_{d \sigma}=0$, $\beta=100$ and the length of the
leads is $N=100$. } \label{one}
\end{figure}

In Fig.\ref{one} we display the contour plot of the TMR in the
parameter space spanned by the bias voltage $U=U_{L}=-U_{R}$ and
the band-spin-splitting $\D \ve$ at inverse temperature
$\beta=100$, $V=0.5$, $\varepsilon_{d \sigma}=0$ and for
$V_{sf}=0,\,0.25,\,0.5$. In the left panel we show the TMR for
$V_{sf}=0$. We can see that despite the dot-leads link is
symmetric, there is a finite region at small bias and intermediate
$\Delta \varepsilon$ in which the TMR $<0$, although rather small
($\approx -0.05$). This is a new scenario for the TMR inversion
and stems from the finite bandwidth of the leads (we recall that
TMR $>0$ for $V_{sf}=0$ in the WBLA). The largest positive value
of the TMR (TMR $\approx 0.7$) occurs for large magnetization and
bias, as expected.

As $V_{sf}$ is increased (central and right panels of
Fig.\ref{one}) the region of positive TMR widens, which is an
\textit{opposite behavior to the one in the } WBLA. However, we
note that the largest value of negative TMR occur at small $U$ and
large $\Delta \varepsilon$ (TMR $\approx -0.2$, see right panel of
Fig.\ref{one}) and that the TMR changes sign abruptly as $U$ is
increased, a feature which is in common with the WBLA. The
positive values of the TMR reduce with respect to the case
$V_{sf}=0$. This property is due to the presence of spin-flip
scatterings which close conducting channels in the P configuration
and open new ones in the AP configuration, thus suppressing the
difference $I_{P}-I_{AP}$.

Finally we have investigated the dependence of the above scenario
on temperature. It is observed that the qualitative picture
described in Fig.\ref{one} survives down to $\beta \approx 10$.
Increasing further the temperature the TMR behaves similarly to
the WBLA.

\section{Summary and conclusions}
\label{secV}

The ultimate goal of future QD-based devices is the possibility to
generate spin-polarized currents, control their spin-coherence
time, and achieve high TMR after the application of high-frequency
signals. This calls for a deep understanding of the time-dependent
responses in these systems.

In this paper we have calculated spin-dependent out-of-equilibrium
properties of lead-QD-lead junctions.  Realistic transient
responses are obtained within the partition-free approach. The
time-dependent current is calculated for QDs connected to
ferromagnetic leads and in the presence of an intra-dot spin flip
interaction. This requires the propagation of a two-component
spinor.

For 1D leads, we evolve exactly a system with a finite number $N$
of sites in each lead. If $N$ is sufficiently large, reliable
time-evolutions are obtained during a time much larger than all
the characteristic time-scales of the infinite
system.\cite{finiteleads1,finiteleads2,finiteleads3} By comparing
our results against the ones obtained with leads of infinite
length,\cite{stefanucci3,stefanucci2} we have verified that our
method is accurate and robust, beside being very easy to
implement.

We have solved analytically the time-dependent problem in the WBLA
and derived a closed formula for the spin-polarized current. Such
formula generalizes the one obtained in the spin-diagonal case and
has a transparent interpretation.\cite{souza,souza2} We stress,
however, that within the WBLA, transitions involving the top or
the bottom of the leads band are not accounted for. The latter may
be relevant to characterize, e.g, the coherent beat oscillations
when the device is attached to narrow-band electrodes.

Furthermore we have shown how to engineer the transient response
of the system to enhance the spin-polarization of the current
through the QD. This is achieved by controlling parameters like,
e.g., the external magnetic field, the transparency of the
contacts, and imposing a pulsed bias of optimal period. It is
shown that by exploiting the synergy between the Pauli-blockade
phenomenon and the resonant-continuum frequency mismatch, one can
achieve an AC spin-polarization two orders of magnitude larger
than the DC one.

We also have employed the Stoner model to describe ferromagnetic
leads and computed the steady-state TMR. We have found a novel
regime of negative TMR, in which the geometry of the tunnel
junction is not required to be asymmetric and a finite intra-dot
spin-flip interaction turns out to be crucial. For any given
$V_{sf}$ there is a critical value of the ferromagnetic
polarization above which the TMR is negative. The magnitude of the
TMR is very sensitive to temperature variations and the TMR
inversion phenomenon disappears as $\beta$ approaches the
damping-time of the system.

We would like to stress that our approach is not limited to 1D
electrodes and can be readily generalized to investigate
multi-terminal devices consisting of several multi-level QDs.
Finally, owning to the fact that the propagation algorithm is
based on a one-particle scheme, it prompts us to include
electron-electron interactions at any mean-field level or within
time-dependent density functional theory.\cite{rg.1984}

\section{Acknowledgements}

E.P. is financially supported by Fondazione Cariplo n. Prot.
0018524. This work is also partially supported by the EU Network
of Excellence NANOQUANTA (NMP4-CT-2004-500198).


\end{document}